\newcommand{\NPA}[3]{Nucl.\ Phys.\ A\ {\bf #1},\ #2 (#3)}
\newcommand{\NPB}[3]{Nucl.\ Phys.\ B\ {\bf #1},\ #2 (#3)}

\newcommand{\PLB}[3]{Phys.\ Lett.\ B\ {\bf #1},\ #2 (#3)}

\newcommand{\PRL}[3]{Phys.\ Rev.\ Lett.\ {\bf #1},\ #2 (#3)}

\newcommand{\PRC}[3]{Phys.\ Rev.\ C\ {\bf #1},\ #2 (#3)}
\newcommand{\PRD}[3]{Phys.\ Rev.\ D\ {\bf #1},\ #2 (#3)}
\newcommand{\JPG}[3]{J.\ Phys.\ G\ {\bf #1},\ #2 (#3)}

\newcommand{\ZPC}[3]{Z.\ Phys.\ C\ {\bf #1},\ #2 (#3)}

\newcommand{\PTP}[3]{Prog.\ Theo.\ Phys.\ {\bf #1},\ #2 (#3)}

\newcommand{\diracslash}[1]{#1\llap{/\kern2pt}}

\newcommand{\be}{\begin{equation}}
\newcommand{\ee}{\end{equation}}
\newcommand{\bea}{\begin{eqnarray}}
\newcommand{\eea}{\end{eqnarray}}
\newcommand{\ba}[1]{\begin{array}{#1}}
\newcommand{\ea}{\end{array}}

\documentclass[prd,aps,floats,nofootinbib,tightenlines,showpacs]{revtex4}
\usepackage{epsfig,graphicx,pstricks}
\usepackage{psfrag}
\usepackage{color}
\usepackage{amsmath}
\usepackage{amsfonts}
\usepackage{amssymb}
\usepackage{textcomp}
\usepackage{multirow}

\begin{document}

\title {Strong CP violation and chiral symmetry breaking in hot and dense quark matter}
\author{Bhaswar Chatterjee}
\email{bhaswar@prl.res.in}
\affiliation{Theory Division, Physical Research Laboratory,
Navrangpura, Ahmedabad 380 009, India}
\author{Hiranmaya Mishra}
\email{hm@prl.res.in}
\affiliation{Theory Division, Physical Research Laboratory,
Navrangpura, Ahmedabad 380 009, India}
\author{Amruta Mishra}
\email{amruta@physics.iitd.ac.in}
\affiliation{Department of Physics, Indian Institute of Technology, New 
Delhi 110016, India}

\date{\today} 

\def\be{\begin{equation}}
\def\ee{\end{equation}}
\def\bearr{\begin{eqnarray}}
\def\eearr{\end{eqnarray}}
\def\zbf#1{{\bf {#1}}}
\def\bfm#1{\mbox{\boldmath $#1$}}
\def\hf{\frac{1}{2}}
\def\sl{\hspace{-0.15cm}/}
\def\omit#1{_{\!\rlap{$\scriptscriptstyle \backslash$}
{\scriptscriptstyle #1}}}
\def\vec#1{\mathchoice
        {\mbox{\boldmath $#1$}}
        {\mbox{\boldmath $#1$}}
        {\mbox{\boldmath $\scriptstyle #1$}}
        {\mbox{\boldmath $\scriptscriptstyle #1$}}
}

\begin{abstract}
We investigate chiral symmetry breaking and strong CP violation effects in the phase diagram of 
strongly interacting matter. We demonstrate the effect of strong CP violating terms on the phase 
structure at finite temperature and densities in a 3-flavor Nambu-Jona-Lasinio (NJL) model 
including the Kobayashi-Maskawa-t'Hooft (KMT) determinant term. This is investigated using an 
explicit structure for the ground state in terms of quark-antiquark condensates for both in the 
scalar and the pseudoscalar channels. CP restoring transition with temperature at zero baryon 
density is found to be a second order transition at $\theta = \pi$ while the same at finite 
chemical potential and small temperature turns out to be a first order transition. Within the 
model, the tri-critical point turns out to be $(T_c,\mu_c)\simeq(273,94)$ MeV at $\theta = \pi$ 
for such a transition.
\end{abstract}

\pacs{12.38.Mh, 12.39.-x, 11.30.Rd, 11.30.Er}

\maketitle

\section{Introduction}
Strong interaction is known to respect space and time reflection symmetry to a very high degree. 
However this is not a direct consequence of laws of quantum chromodynamics (QCD), which, in 
principle permit a parity violating term or the so called $\theta$-term given as

\begin{equation}
{\cal L}_\theta=\frac{\theta}{64\pi^2} g^2 F_{\mu\nu}^a\tilde F^{a\mu\nu}.
\label{lth}
\end{equation}

In the above, $F_{\mu\nu}^a$ is the gluon field strength and $\tilde F^{\mu\nu}$ being its dual. 
This term while being consistent with Lorentz invariance and gauge invariance, it violates charge 
conjugation and parity unless $\theta=0$ mod $\pi$. However, CP symmetry conserving nature of QCD 
has been established by precise experiments that sets limit on the intrinsic electric dipole moment 
of neutron. The current experimental limit on this leads to limit on the coefficient of the CP 
violating term of the QCD Lagrangian as $\theta < 0.7\times 10^{-11}$ \cite{endm}. This smallness 
of the CP violation term or its complete absence is not understood completely though a possible 
explanation is given in terms of spontaneous breaking of a new symmetry : the Peccei-Quinn symmetry 
\cite{pecceiquinn} which could give rise to axions. For zero temperature and and zero density, 
spontaneous parity violation does not arise for $\theta=0$ by the well known Vafa-Witten theorem 
\cite{vafawit}. On the other hand for $\theta=\pi$ there could be spontaneous CP violation by the 
so called Dashen phenomena \cite{dashen}. Because of the nonperturbative nature of this $\theta$ 
term in QCD, this problem has been studied extensively in low energy effective theories like chiral 
perturbation theory \cite{chpt}, linear sigma model \cite{fragalsm} as well as Nambu-Jona-Lasinio 
(NJL) model and its different extensions \cite{cpnjl,bbone,bbtwo,sakai}.

Even if CP is not violated for QCD vacuum, it is possible that it can be violated for QCD matter at 
finite temperature or density. It has been proposed that hot matter produced in heavy ion collision 
experiments can give rise to domains of meta stable states that violate CP \cite{dimacp}. 
Experimental signatures for the existence of local CP violation has been based on charge separation 
of hadronic matter due to the strong magnetic field produced in heavy ion collision experiments by 
a mechanism called chiral magnetic effect (CME) \cite{cme}. This mechanism may explain the charge 
separation in the recent STAR results \cite{starexp}.

In the present work we focus our attention on how chiral transition is affected when there is CP 
violating term in the Lagrangian. For this purpose, we adopt the 3-flavor NJL model as an effective 
theory for chiral symmetry breaking in strong interaction \cite{klevansky,rehberg}. The CP violating 
parameter $\theta$ is included in the Kobayashi-Maskawa-t'Hooft (KMT) determinant term. In this 
context we note that the two flavor scenario for spontaneous CP violation for $\theta=\pi$ has been 
studied in this model \cite{bbone}. This has been further extended to study the restoration of CP at 
finite temperature \cite{bbtwo}. The effect of the theta vacuum on the deconfinement and chiral 
transition has also been analyzed within a two flavor NJL model with Polyakov loop \cite{sakai}. 

We organize the present work as follows. In the next section we shall discuss the 3-flavor NJL model 
with a CP violating term. We consider a variational ground state with quark-antiquark pairs that is 
related to chiral symmetry breaking. The ansatz functions are to be determined through minimization 
of the thermodynamic potential. The ansatz is general enough to include both scalar as well as 
pseudoscaler condensates. As we shall see the pseudoscalar condensates develop for non zero values 
of $\theta$ in the KMT determinant term. In section III we discuss the resulting phase diagram at 
finite temperature as well as finite density for different values of the CP violating parameter in 
the Lagrangian. In Section IV we summarize our results and give a possible outlook.

\section{NJL model with CP violation and an ansatz for the ground state}
To describe the chiral phase structure of strong interactions including the CP violating effects, 
we use the 3-flavor NJL model along with the flavor mixing determinant term. The Lagrangian is 
given by

\begin{equation}
{\cal L} = \bar\psi\left(i\partial\sl - m\right)\psi
+ G\sum_{A=0}^8\left[(\bar\psi\lambda^A\psi)^2 + (\bar\psi i\gamma^5\lambda^A\psi)^2\right]
- K\left[e^{i\theta}det\lbrace\bar\psi(1+\gamma^5)\psi\rbrace +
e^{-i\theta}det\lbrace\bar\psi(1-\gamma^5)\psi\rbrace\right],
\label{lag3fl}
\end{equation}

\noindent
where $\psi ^{i,a}$ denotes a quark field with color `$a$' $(a=r,g,b)$, and flavor `$i$' 
$(i=u,d,s)$, indices. The matrix of current quark masses is given by $\hat m$ = 
diag$_f(m_u,m_d,m_s)$ in the flavor space. We shall assume in the present investigation, isospin 
symmetry with $m_u$ = $m_d$. In Eq.(\ref{lag3fl}), $\lambda^A$, $A=1,\cdots 8$ denote the 
Gell-Mann matrices acting in the flavor space and 
$\lambda^0 = \sqrt{\frac{2}{3}}\,1\hspace{-1.5mm}1_f$, where $\,1\hspace{-1.5mm}1_f$ is the unit 
matrix in the flavor space. The four point interaction term $\sim G$ is symmetric under 
$SU(3)_V\times SU(3)_A\times U(1)_V\times U(1)_A$. The determinant term $\sim K$, which generates 
a six point interaction for the case of three flavors, breaks $U(1)_A$ symmetry for vanishing 
$\theta$ values. The effect of topological term of Eq.(\ref{lth}) is simulated by the determinant 
term of Eq.(\ref{lag3fl}) in the quark sector.

We shall next consider an ansatz for the ground state with quark-antiquark condensates which 
includes both the scalar as well as CP violating pseudoscalar channel. To make notations clear, 
we first write down the field operator expansion for the quark fields as given in 
\cite{hmspmnjl,amspm}, 

\begin{equation}
\psi (\zbf x,t=0 )\equiv \frac{1}{(2\pi)^{3/2}}\int \tilde\psi(\zbf k)
e^{i\zbf k\cdot\zbf x}d\zbf k 
=\frac{1}{(2\pi)^{3/2}}\int \left[U_0(\zbf k)q^0(\zbf k )
+V_0(-\zbf k)\tilde q^0(-\zbf k )\right]e^{i\zbf k\cdot \zbf x}d \zbf k,
\label{psiexp}
\end{equation}

\noindent
where $U^0(\zbf k)$ and $V^0(-\zbf k)$ are the four component spinors which can be explicitly 
written as,

\begin{eqnarray}
U_0(\zbf k )=&&\left(\begin{array}{c}\cos(\frac{\chi^0}{2})\\
\zbf \sigma \cdot \hat k \sin(\frac{\chi^0}{2})
\end{array}\right)\;\;\;\;\;\;\;\;and\;\;\;\;\;\;\;\;
V_0(-\zbf k )=
\left( \begin{array}{c} -\zbf \sigma \cdot \hat k  \sin(\frac{\chi^0}{2})
\\ \cos(\frac{\chi^0}{2})\end{array}
\right).
\label{uv0}
\end{eqnarray}

The superscript `$0$' indicates that the operators $q^0$ and $\tilde q^0$ are the two component 
operators for the quark annihilation and antiquark creation corresponding to the the perturbative 
or the chiral vacuum $|0\rangle$. Here we have suppressed the color and flavor indices of the quark 
field operators. The function $\chi^0(\zbf k)$ in the spinors in Eq.(\ref{uv0}) are given as
$\cot{\chi_i^0}=m_i/|\zbf k|$, for free massive fermion fields, $i$ being the flavor index. For 
massless fields $\chi^0(|\zbf k|)=\pi/2$.

We next consider an ansatz of  the ground state at zero temperature as

\begin{equation}
|\Omega\rangle=U_q|0\rangle,
\end{equation}

where, $U_q=U_{qI}U_{qII}$ is an unitary operator. $U_{qI}$ and $U_{qII}$ are unitary operators 
described in terms of quark-antiquark creation and annihilation operators. Explicitly they are 
given as

\begin{equation}
U_{qI}=\exp\left(\int d\zbf k q_r^0(\zbf k)^\dagger (\bfm\sigma\cdot\hat{\zbf k})_{rs}
f(k)\tilde q_{s}^0(-\zbf k)-h.c\right)
\label{uqI}
\end{equation}

and

\begin{equation}
U_{qII}=\exp\left(\int d\zbf k q_r(\zbf k)^\dagger
r  g(k)\tilde q_{-r}(-\zbf k)-h.c\right)
\end{equation}

where $f(k)$ and $g(k)$ are the ansatz functions which are to be determined later from the 
extremization of the thermodynamic potential.

Finally, to include the effect of temperature and baryon density, we use the techniques of 
thermofield dynamics (TFD) which is quite convenient while dealing with operators and states 
\cite{tfd,amph4}. Here, the statistical average of an operator is given as an expectation value 
over a `thermal vacuum'. The methodology of TFD involves the doubling of the Hilbert space 
\cite{tfd}. Explicitly, the `thermal vacuum' is constructed from the ground state at zero 
temperature and density through a thermal Bogoliubov transformation given as

\begin{equation}
|\Omega(\beta,\mu)\rangle = {\cal U}_F|\Omega\rangle = 
e^{{\cal B}(\beta,\mu)^\dagger-{\cal B}(\beta,\mu)}|\Omega\rangle
\label{ubt}
\end{equation}

with,

\begin{equation}
{\cal B}^\dagger(\beta,\mu)=\int \Big [ \theta_-(\zbf k, \beta,\mu)
q^\prime (\zbf k)^\dagger \underline q^{\prime} (-\zbf k)^\dagger +
\theta_+(\zbf k, \beta,\mu) \tilde q^\prime (\zbf k)
\underline { \tilde q}^{\prime} (-\zbf k)
\Big ] d\zbf k.
\label{bth}
\end{equation}

In Eq.(\ref{bth}) the ansatz functions $\theta_{\pm}(\zbf k,\beta,\mu)$ will be related to the 
quark and the antiquark thermal distributions respectively and the underlined operators are the 
operators in the extended Hilbert space associated with thermal doubling in TFD method.

In the following section we shall compute the thermodynamic potential which will involve 
calculating thermal average of different operators which is given by the expectation values of the 
corresponding operator with respect to the state  given in Eq.(\ref{ubt}). This can be evaluated 
directly by realizing that the state $|\Omega(\beta,\mu)\rangle$ is obtained from the state 
$|0\rangle$ by successive Bogoliubov transformation. Thus e.g. we have

\begin{equation}
\langle \Omega(\beta\mu)|
\psi^{\dagger i a}_\alpha (\zbf x)\psi^{j b}_\beta (\zbf y)
|\rangle\Omega(\beta\mu)\rangle=\delta^{ij}\delta^{ab}\int\frac{d\zbf k}{(2\pi)^3}
e^{-i\zbf k\cdot(\zbf x-\zbf y)}\Lambda^i(\zbf k,\beta,\mu)_{\beta\alpha}
\label{master}
\end{equation}

where,

\begin{equation}
\Lambda^i(\zbf k,\beta,\mu) = \frac{1}{2}\left[(\cos^2\theta^i_++\sin^2\theta_-^i)
+ (\sin^2\theta^i_--\sin^2\theta_+^i)(\gamma^0\cos\phi^i\cos 2g^i+\zbf\alpha\cdot\hat
{\zbf  k}
\sin\phi^i\cos 2g^i-i\gamma^0\gamma^5\sin 2g^i\right]
\end{equation}

where, we have introduced a new function $\phi^i(\zbf k)=\chi_0+2 f^i(\zbf k)$ in terms of the 
condensate function $f^i(\zbf k)$ of Eq.(\ref{uqI}). From Eq.(\ref{master}), it is easy to 
calculate the scalar and pseudoscalar condensates. In terms of the ansatz functions 
$\phi^i(\zbf k)$ and $g^i(\zbf k)$ the scalar and pseudoscalar condensates for the i-th flavor 
can be respectively written as 

\begin{eqnarray}
\langle\bar\psi\psi\rangle_i &=& 
-\frac{2N_c}{(2\pi)^3}\int d\zbf k\cos \phi^i\cos{2g^i}(1-n^i_- - n^i_+) \equiv -I_s^i\\
\label{scalar}
\langle\bar\psi\gamma_5\psi\rangle_i &=& 
-i\frac{2N_c}{(2\pi)^3}\int d\zbf k\sin{2g^i}(1-n^i_- - n^i_+) \equiv
-iI_p^i
\label{pseudo}
\end{eqnarray}

where $n^i_\mp=\sin^2\theta^i_\mp$. Thus a non vanishing $I_s^i$ will imply chiral symmetry 
breaking phase while a non vanishing $I_p^i$ or equivalently $g^i(\zbf k)$ will indicate CP 
violating phase. The condensate functions $\phi^i(\zbf k)$, $g^i(\zbf k)$ as well as the thermal 
functions $\theta^i_\mp(\zbf k,\beta,\mu)$ shall be determined by extremization of the 
thermodynamic potential with respect to the respective functions. We shall carry out these 
extremization in the following section.

\section{Evaluation of thermodynamic potential and gap equations }
\label{evaluation}

As mentioned we shall be considering the chiral phase structure in presence of the CP violating 
terms within the frame work of Nambu-Jona-Lasinio model of Eq.(\ref{lag3fl}). The energy density 
is given by the expectation value of the Hamiltonian corresponding to the Lagrangian given in 
Eq.(\ref{lag3fl}) with respect to the thermal ansatz of Eq.(\ref{ubt}). The energy density can be 
written as

\begin{equation}
\epsilon = T + V = T + V_S + V_D
\label{entot}
\end{equation}

where $T$ is the expectation value of the kinetic term in Eq.(\ref{lag3fl}) and using 
Eq.(\ref{master}), is given as

\begin{equation}
T = \langle\psi^\dag(-i\vec\alpha\cdot\vec\nabla+\beta m)\psi\rangle 
  = -\frac{2N_c}{(2\pi)^3}\sum_i\int{d\zbf k(m^i\cos \phi^i + |\zbf k|\sin \phi^i)
    \cos{2g^i}(1-n^i_--n^i_+)}.
\label{kinetic}
\end{equation}

$V_S$ the contribution from the four point interaction term in Eq.(\ref{lag3fl}) to the energy 
density and using Eq.(\ref{master}), this is given as

\begin{equation}
V_S = -G\langle\sum_{A=0}^8\left[(\bar\psi\lambda^A\psi)^2 + 
(\bar\psi i\gamma_5\lambda^A\psi)^2\right]\rangle
 = -2G\sum_i\left[{I_s^i}^2 + {I_p^i}^2\right].
\label{vs}
\end{equation}

Finally, $V_D$ denotes the contribution from the determinant interaction term in Eq.(\ref{lag3fl}) and 
using Eq.(\ref{master}), is given as

\begin{eqnarray}
V_D &=& K\langle e^{i\theta}det\lbrace\bar\psi(1+\gamma_5)\psi\rbrace + 
e^{-i\theta}det\lbrace\bar\psi(1-\gamma_5)\psi\rbrace\rangle \nonumber\\
 &=& 2K\left[\cos\theta\left\lbrace-\prod_{i=1}^3I_s^i +
 \frac{1}{2}|\epsilon_{ijk}|I_s^i I_p^j I_p^k\right\rbrace
 + \sin\theta\left\lbrace-\prod_{i=1}^3I_p^i +
 \frac{1}{2}|\epsilon_{ijk}|I_s^i I_s^j I_p^k\right\rbrace\right].
\label{vd}
\end{eqnarray}

Now, the thermodynamic potential is given as

\begin{equation}
\Omega = \epsilon  -\mu \rho- \frac{S}{\beta},
\label{omega1}
\end{equation}

where $\mu$ is the quark chemical potential corresponding to the quark number density $\rho$ 
given as

\begin{equation}
\rho = \sum_{i=u,d,s}\langle\psi^\dag\psi\rangle_i =
\frac{2N_c}{(2\pi)^3}\sum_{i=u,d,s}\int{d\zbf k(1-\sin^2\theta^i_+ + \sin^2\theta^i_-)}.
\label{numden}
\end{equation}

Finally, $S$ is the entropy density given as

\begin{equation}
S =\frac{2N_c}{(2\pi)^3}\sum_{i=u,d,s}
\int d\zbf k\left(\cos^2\theta_-^i\ln \cos^2\theta_-^i+
\sin^2\theta_-^i\ln\sin^2\theta_-^i
+\cos^2\theta_+^i\ln \cos^2\theta_+^i
+ \sin^2\theta_+^i\ln\sin^2\theta_+
\right).
\label{sq}
\end{equation}

Thus the thermodynamic potential given in Eq.(\ref{omega1}) is known in terms of the ansatz 
functions of Eq.(\ref{ubt}). Extremizing the thermodynamic potential with respect to 
$\phi^i(\zbf k)$ and $g^i(\zbf k)$ respectively leads to

\begin{eqnarray}
\tan \phi^i = \frac{|\zbf k|}{M_s^i}\;\;\;\;\;\;\;\;and\;\;\;\;\;\;\;\;
\tan{2g^i} = \frac{M_p^i}{\sqrt{{M_p^i}^2 + |\zbf k|^2}}.
\end{eqnarray}

where $M_s^i$ and $M_p^i$ are respectively the contributions to the constituent quark mass 
(for i-th flavor) from the scalar and pseudoscalar condensates and they are given by,

\begin{eqnarray}
M_s^i &=& m^i + 4GI_s^i + K|\epsilon_{ijk}|\lbrace\cos\theta(I_s^j I_s^k - I_p^j I_p^k) - 
\sin\theta(I_s^j I_p^k + I_p^j I_s^k)\rbrace \\
\label{scmass}
M_p^i &=& 4G I_p^i - K|\epsilon_{ijk}|\lbrace\cos\theta(I_s^j I_p^k + I_p^j I_s^k) - 
\sin\theta(I_p^j I_p^k - I_s^j I_s^k)\rbrace.
\label{psmass}
\end{eqnarray}

Finally extremizing the thermodynamic potential with respect to the thermal
distribution functions $\theta^i_{\mp}$ leads to

\begin{equation}
\sin^2\theta^i_\pm=\frac{1}{\exp(\omega^i\mp\mu^i)+1},
\label{sth}
\end{equation}

where, $\omega^i(\zbf k)=\sqrt{\zbf k^2+({M_s^i}^2+{M_p^i}^2)}$. Thus we can see that the 
constituent quark masses get contribution from both the scalar and pseudoscalar condensates.

Substituting the extremized solution for the condensate functions $\tan\phi^i$ and $\tan{2g^i}$ 
in Eq.(\ref{scalar}) and Eq.(\ref{pseudo}), we have the self consistent equations for the scalar 
and pseudoscalar condensates,

\begin{eqnarray}
I_s^i &\equiv& -\langle\bar\psi\psi\rangle_i = 
\frac{2N_c}{(2\pi)^3}\int{d\zbf k\left(1-n_-^i-n_+^i\right)\frac{M_s^i}{\omega^i}}
\label{isi}\\
I_p^i &\equiv& i\langle\bar\psi\gamma_5\psi\rangle_i = 
\frac{2N_c}{(2\pi)^3}\int{d\zbf k\left(1-n_-^i-n_+^i\right)\frac{M_p^i}{\omega^i}}.
\label{ipi}
\end{eqnarray}

Thus with the scalar and the pseudoscalar condensates given as above, 
Eq.s(\ref{scmass}-\ref{psmass}) are actually coupled self-consistent equations for $M_s^i$ and 
$M_p^i$.

Substituting the extremized solutions for the condensate functions and using the gap equations 
Eq.(\ref{scmass}-\ref{psmass}) in Eq.(\ref{omega1}), the thermodynamic potential becomes

\begin{eqnarray}
\Omega &=& -\frac{2N_c}{(2\pi)^3}\sum_i\int{d\zbf k(\omega^i-|\zbf k|)} +
 2G_s\sum_i\left[{I_s^i}^2 - {I_p^i}^2\right] + \sum_i M_p^i I_p^i\nonumber\\
 &&+ 4K\cos\theta\prod_{i=1}^3I_s^i - 2K\sin\theta\left[\prod_{i=1}^3I_p^1 +
 \frac{1}{2}|\epsilon_{ijk}|I_s^i I_s^j I_p^k\right] \nonumber \\
 &&- \frac{2N_c}{\beta(2\pi)^3}\sum_i\int{d\vec k\left[\ln{\lbrace 1+e^{-\beta(\omega^i-\mu^i)}\rbrace} +
 \ln{\lbrace 1+e^{-\beta(\omega^i+\mu^i)}\rbrace}\right]}.
\label{thermpot}
\end{eqnarray}

In the above we have subtracted the perturbative vacuum energy density contribution. Let us note 
that, the effective potential has been calculated using an explicit ansatz for the condensate and 
not evaluating the effective potential at a mean field level after performing a chiral 
transformation for the quarks so as to remove it from the determinant term as it has been computed 
in Ref.\cite{bbone,sakai}.

Eq.(\ref{thermpot}) for the thermodynamic potential and the gap equations for the scalar and 
pseudoscalar masses i.e. Eq.(\ref{scmass}-\ref{psmass}) shall be the focus of our numerical analysis 
which we do in the following section.

\section{results and discussions}

For numerical calculations, we have taken the values of the parameters of the NJL model as follows. 
The coupling constant $G_s$ has the dimension of $[{\rm Mass}]^{-2}$ while the six fermion coupling 
$K$ has a dimension $[{\rm Mass}]^{-5}$. To regularize the divergent integrals we use a sharp 
cut-off, $\Lambda$ in 3-momentum space. Thus we have five parameters in total, namely the current 
quark masses for the non strange and strange quarks, $m_q$ and $m_s$, the two couplings $G_s$, $K$ 
and the three-momentum cutoff $\Lambda$. We have chosen here $\Lambda=0.6023$ GeV, 
$G_s\Lambda^2=1.835$, $K\Lambda^5=12.36$, $m^q=5.5$ MeV and $m^s=0.1407$ GeV as has been used in 
Ref.\cite{rehberg}. After choosing $m^q=5.5$ MeV, the remaining four parameters are fixed by fitting 
to the pion decay constant and the masses of pion, kaon and $\eta'$. With this set of parameters the 
mass of $\eta$ is underestimated by about six percent and the constituent masses of the light quarks 
turn out to be $M^{u,d}=0.368$ GeV for u-d quarks, while the same for strange quark turns out as 
$M^s=0.549$ GeV, at zero temperature and zero density.

For a given temperature and the chemical potential, we first solve the coupled self consistent gap 
equations Eq.s(\ref{scmass}-\ref{psmass}) with the parameters of the model as above. Since we have 
assumed isospin symmetry and have $m^u=m^d$, these are actually four coupled equations: two for the 
scalar condensates related to the two masses $M_s^u=M_s^d$, $M_s^s$ and two equations for the 
pseudoscalar condensate related to the corresponding mass parameters $M_p^u=M_p^d$, $M_p^s$. The 
solutions to these equations are then substituted in Eq.(\ref{thermpot}) and checked regarding 
minimum of the thermodynamic potential. If there are more solutions to the gap equation, the one 
with the minimum thermodynamic potential is chosen.

Let us first discuss the ground state structure at zero temperature and zero density. In 
Fig.(1-a) we show the theta dependence of contributions to the mass of up quark from the scalar as 
well as the pseudoscalar condensates. As is clearly seen as $\theta$ increases the condensates in 
the two channels behave in a complimentary manner. While the magnitude of scalar condensates 
decreases with $\theta$ (till $\theta=\pi$), the magnitude of the pseudoscalar condensate increases 
so that the total constituent quark mass $M=\sqrt{{M_s}^2+{M_p}^2}$ remains almost the same. 
Spontaneous CP violation is clearly seen for $\theta=\pi$ with two degenerate solutions for $M_p^u$ 
differing by a sign. In Fig.(1-b) we show the effective potential as calculated above as a function 
of $\theta$. The effective potential is normalized with respect to the same at $\theta=0$. The 
minimum of the potential is at $\theta=0$ which is consistent with the Vafa-Witten theorem and has 
a cusp at $\theta=\pi$ which has also been observed in 2-flavor NJL model \cite{bbone}.

\begin{figure}[t]
\vspace{-0.4cm}
\begin{center}
\begin{tabular}{c c}
\includegraphics[width=9cm,height=9cm]{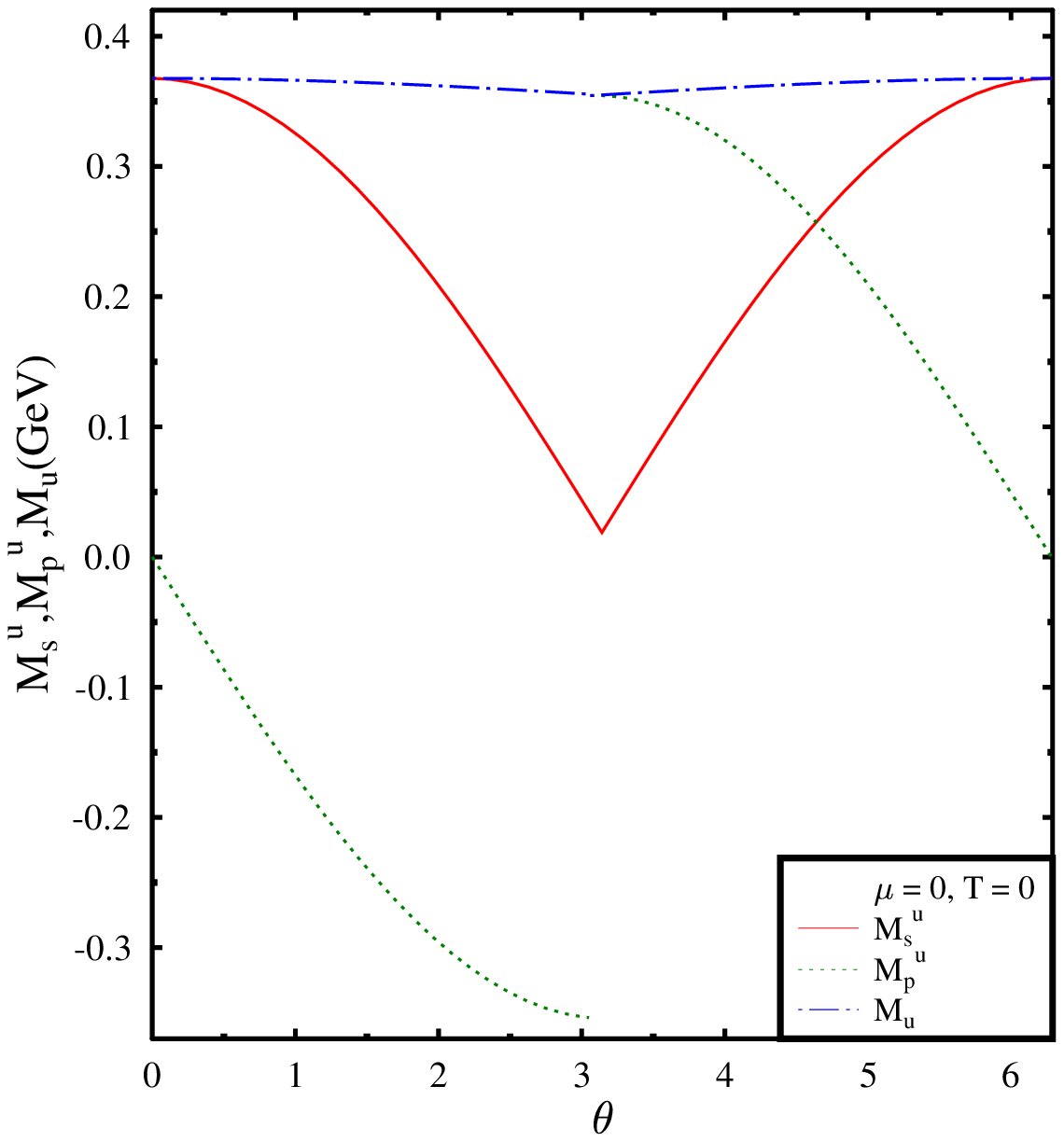}&
\includegraphics[width=9cm,height=9cm]{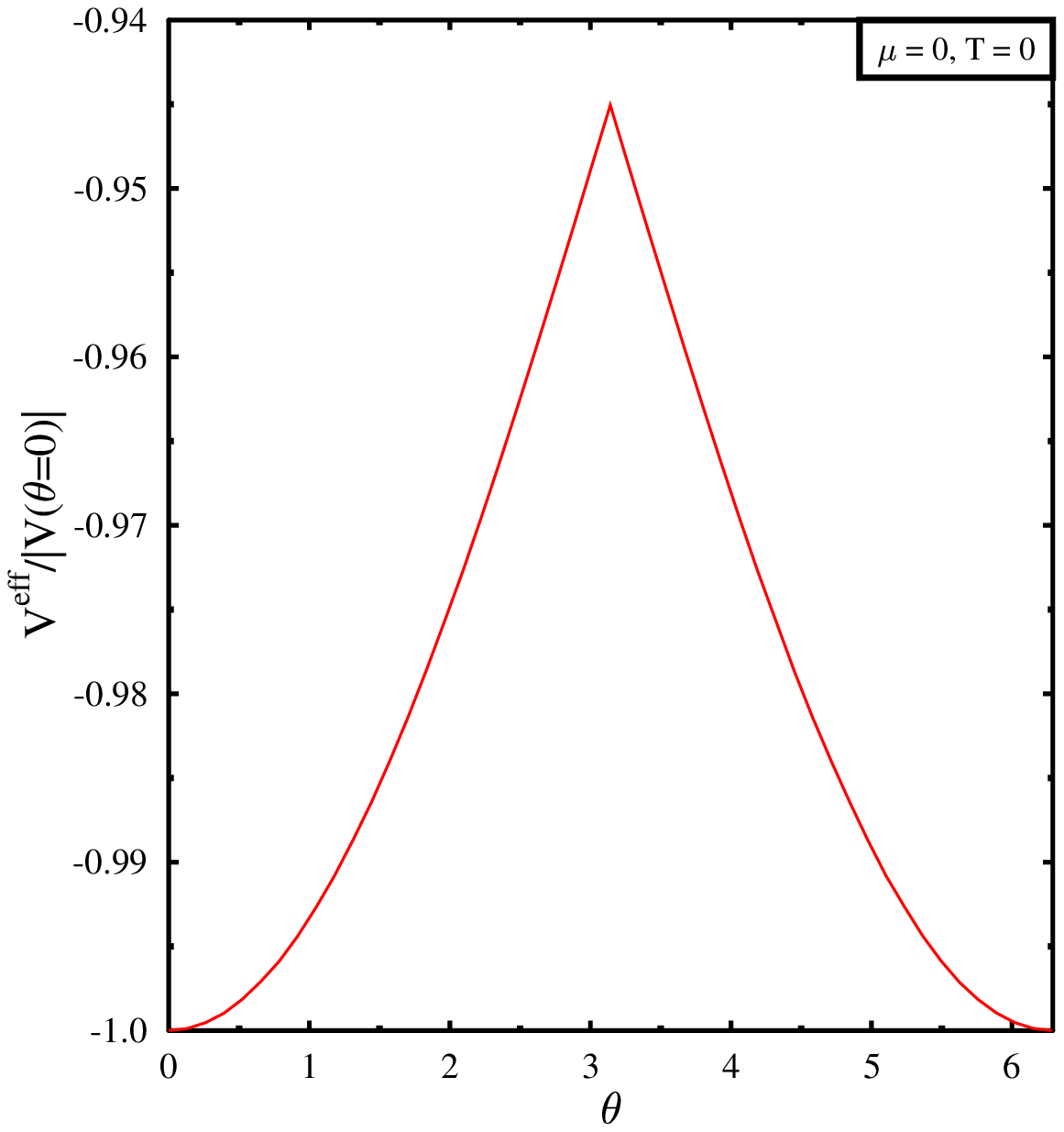}\\
Fig. 1-a & Fig. 1-b
\end{tabular}
\end{center}
\caption{  $\theta$-dependence of the condensates (Fig.1-a) and the effective potential
at zero temperature and zero baryon density (Fig.1-b).}
\label{fig1}
\end{figure}

\begin{figure}[t]
\vspace{-0.4cm}
\begin{center}
\begin{tabular}{c c c}
\includegraphics[width=6cm,height=6cm]{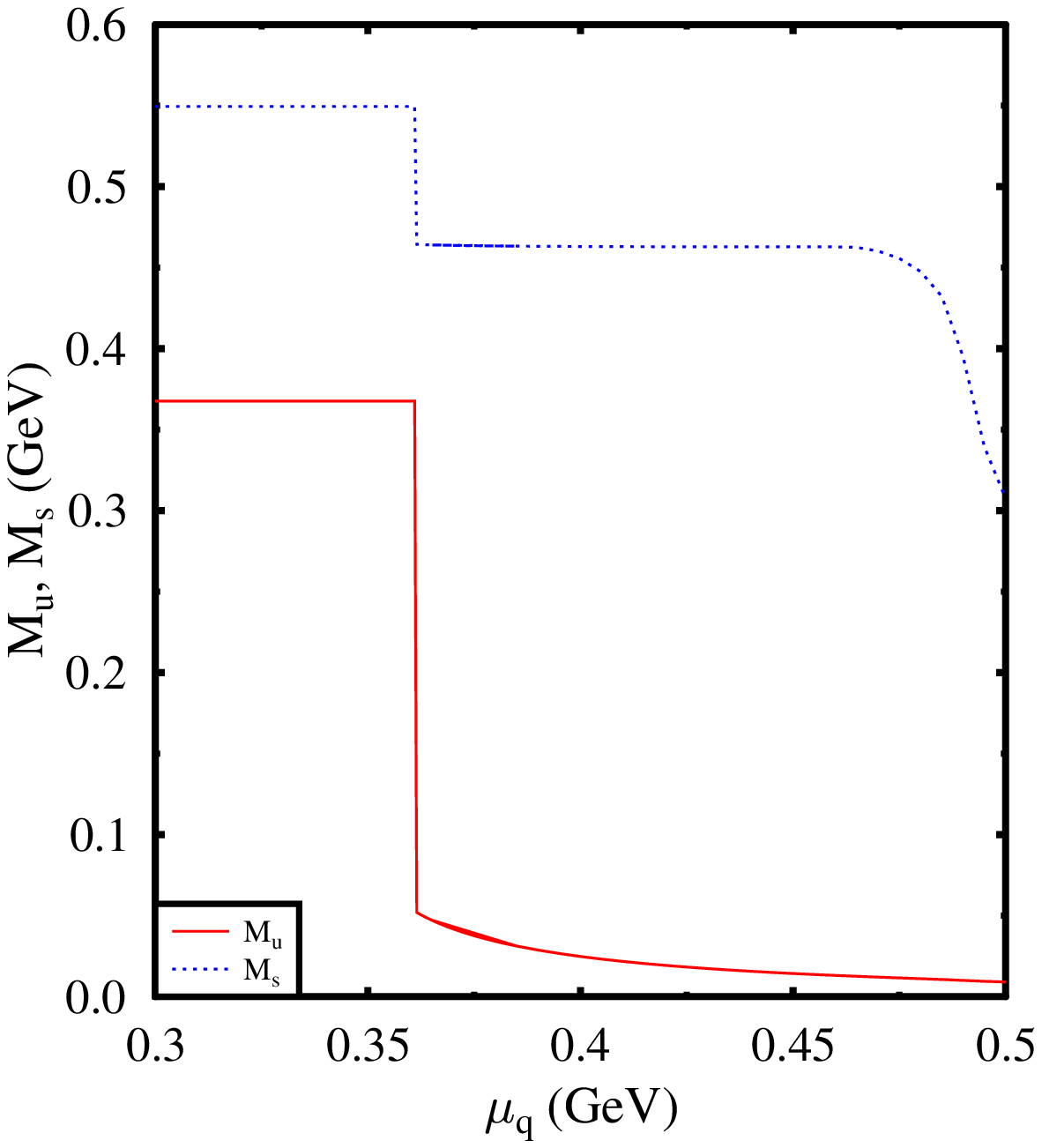}&
\includegraphics[width=6cm,height=6cm]{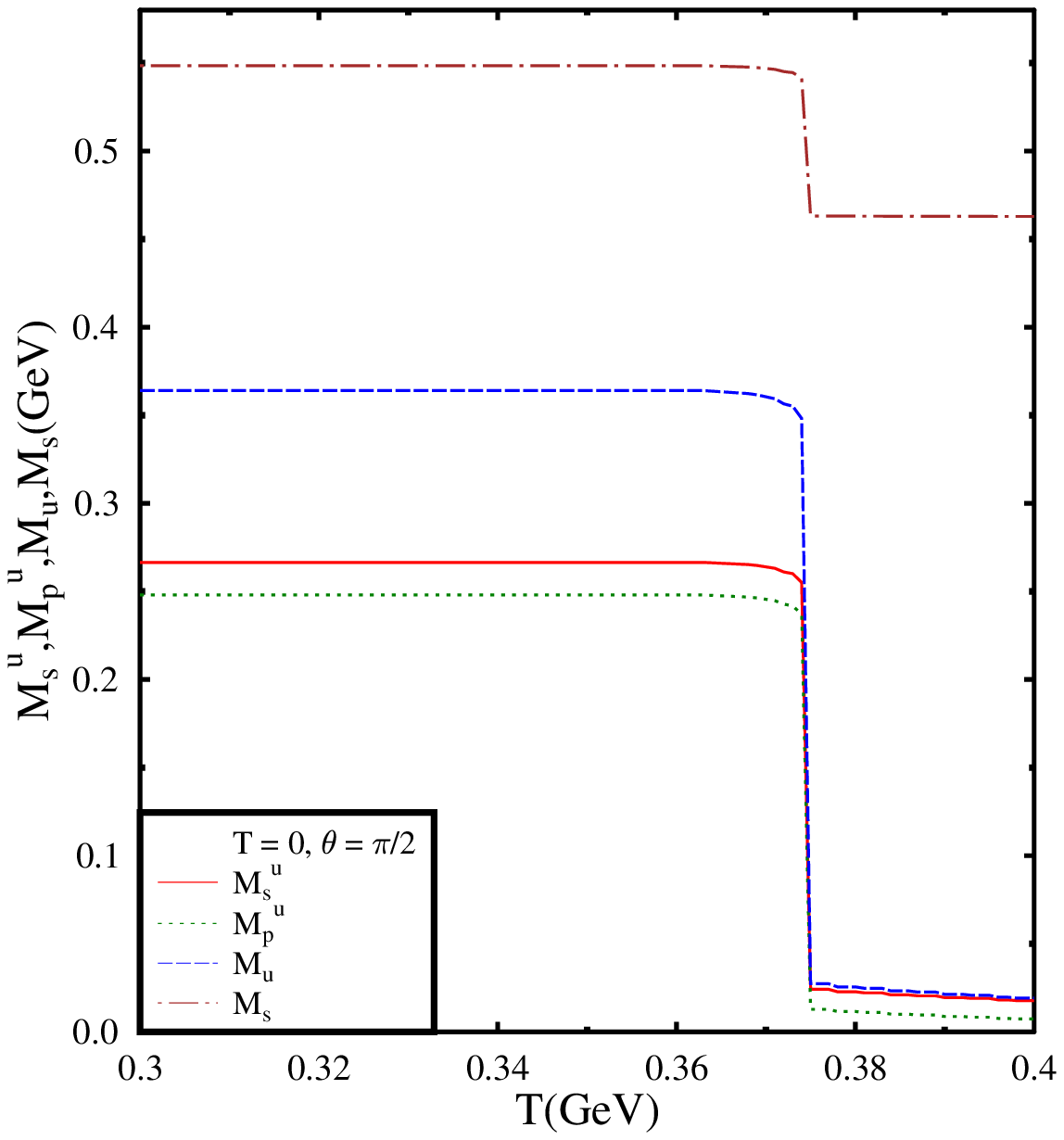}&
\includegraphics[width=6cm,height=6cm]{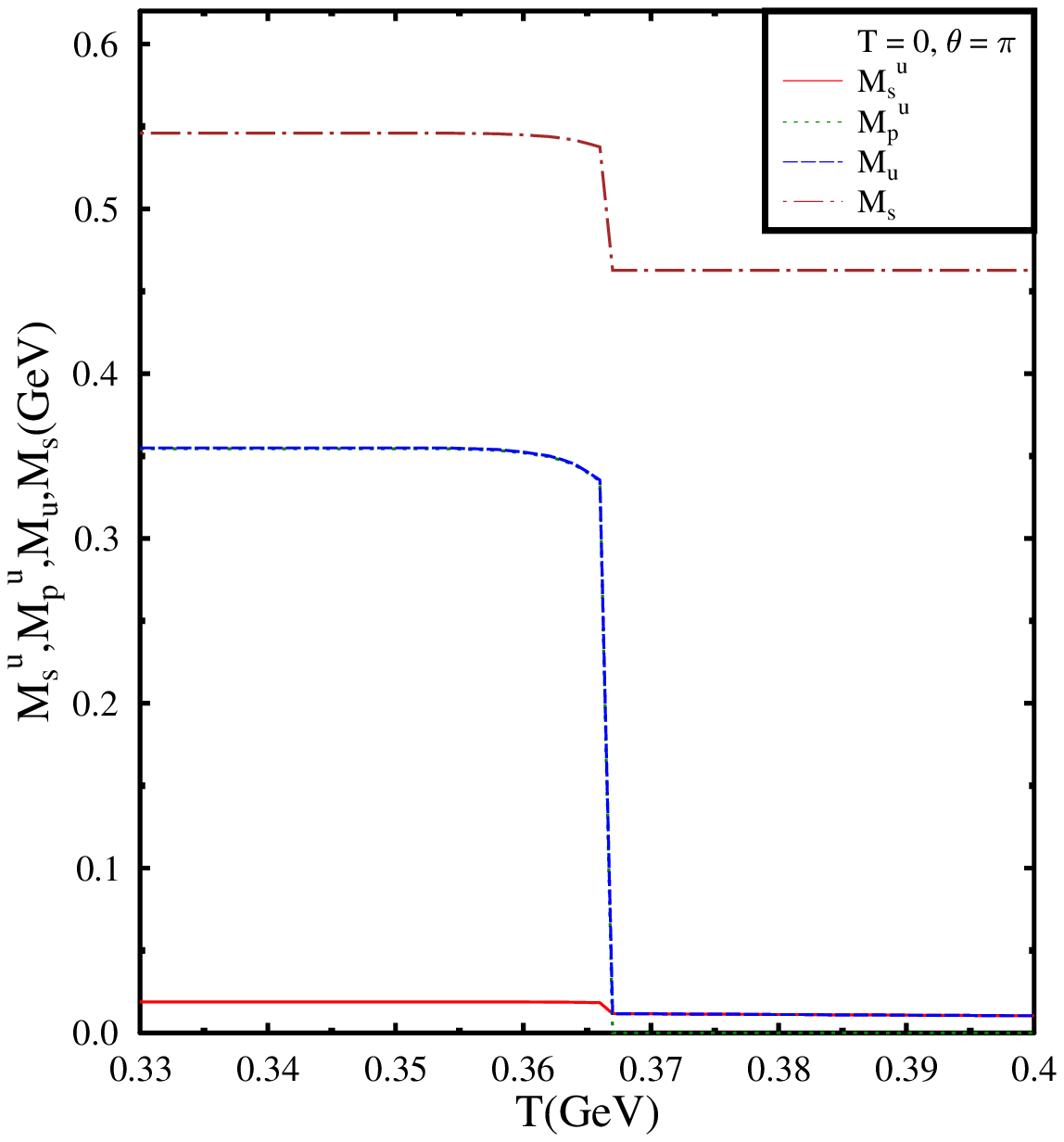}\\
Fig. 2-a & Fig. 2-b & Fig. 2-c
\end{tabular}
\end{center}
\caption{Quark masses as a function of quark chemical potential at zero temperature 
for $\theta=0$ (Fig.2-a), $\theta=\pi/2$ (Fig.2-b) and $\theta=\pi$ (Fig.2-c). The pseudoscalar 
contribution for strange quarks is zero in this range of chemical potential.}
\label{fig2}
\end{figure}

Next we consider the effect of nonzero density and temperature. In Fig.(\ref{fig2}), we show the 
variation of masses of the quarks with chemical potential at zero temperature. In Fig.(2-a) we 
show the variation of masses for the case $\theta=0$. In this case the pseudoscalar condensates 
vanish and the contribution to the masses of the quarks are from the scalar condensates only. The 
(approximate) first order chiral transition takes place at $\mu\sim361$ MeV for u and d quarks 
with their masses decreasing discontinuously to about $M^{u,d}\sim52$ MeV from their vacuum value 
of $M^{u,d}=368$ MeV. Because of the flavor mixing KMT term, this decrease is reflected also in the 
decrease of the strange quark mass to $M^s=464$ MeV from its vacuum value of $M^s=549$ MeV. This 
result is similar to the results obtained in the context of color superconductivity in NJL model 
with a determinant term \cite{hmcsc} and in the context of chiral symmetry breaking in a similar 
model \cite{bccsb}. Similarly in Fig.(2-b) we show the variation of masses of up and strange quark 
as well as the variation of the contributions from scalar and pseudoscalar condensates to the 
constituent mass of up quark for $\theta=\pi/2$. Here the critical chemical potential for chiral 
transition is $\mu_c\sim 375$ MeV where the mass contributions from scalar and pseudoscalar 
condensate become 24 MeV and 12 MeV respectively from their vacuum values of 266 MeV and 248 MeV. 
The total mass for the u and d quarks become 27 MeV from its vacuum value of 364 MeV. On the 
other hand the contribution to the strange mass from the pseudoscalar condensate is negligible 
($\sim 12$ MeV) compared to the contribution from the scalar condensate ($\sim 548$ MeV). Because 
of flavor mixing again, strange quark mass also decrease to 463 MeV at $\mu_c=375$ MeV. For 
$\theta=\pi$ the scalar condensate almost vanishes but for the nonzero current quark masses while 
the contribution to the constituent quark mass arises from the pseudoscalar condensate as shown in 
Fig.(2-c). As the quark chemical potential is increased there is a first order transition at 
$\mu_c\sim 368$ MeV. At $\mu_c$ the pseudoscalar condensate vanishes and the contribution to quark 
mass arises solely from the scalar condensate which is non vanishing because of the nonzero current 
quark masses.

Next we discuss the condensate variations with temperature. For $\theta=0$, at zero baryon density 
the chiral crossover transition takes place for temperature about 200 MeV as may be seen in 
Fig.(3-a). As $\theta$ increases, the pseudoscalar condensates starts becoming nonzero and increase 
with $\theta$. For $\theta=\pi/2$, masses arising from both type of condensates are shown in 
Fig.(3-b). Here, both the scalar and pseudoscalar masses show a crossover transition as temperature 
is increased. In Fig.(3-c), we show the behavior of the masses for $\theta=\pi$. The transition 
for the pseudoscalar mass becomes a second order transition at $\theta=\pi$ instead of a crossover 
which was the nature of transition for lower $\theta$. This feature is elaborated in 
Fig.(\ref{fig4}) where $\theta$ dependence of the nature of transition of scalar and pseudoscalar 
condensates with temperature is shown for zero baryon density. Fig.(4-a) shows the $\theta$ 
dependence of the transitions for the scalar condensate. The transition is always a crossover for 
scalar condensate. Fig.(4-b) shows the transitions of the pseudoscalar condensates for different 
$\theta$. We can see that the transition is a second order transition for $\theta=\pi$ whereas it 
is a crossover for other values of $\theta$. Similar kind of results have been obtained in sigma 
model calculations \cite{fragalsm} but there the transition at $\theta=\pi$ is a first order 
transition instead of a second order transition. The CP restoring transition temperature for zero 
baryon density case turns out to be 192 MeV. However, the total constituent mass is nonzero as the 
scalar condensate is non vanishing again because of nonzero current quark masses. This high 
temperature restoration of CP is expected as the instanton effects responsible for CP violating 
phase become suppressed exponentially at high temperature \cite{grosspisarski}.

\begin{figure}[htbp]
\vspace{-0.4cm}
\begin{center}
\begin{tabular}{c c c}
\includegraphics[width=6cm,height=6cm]{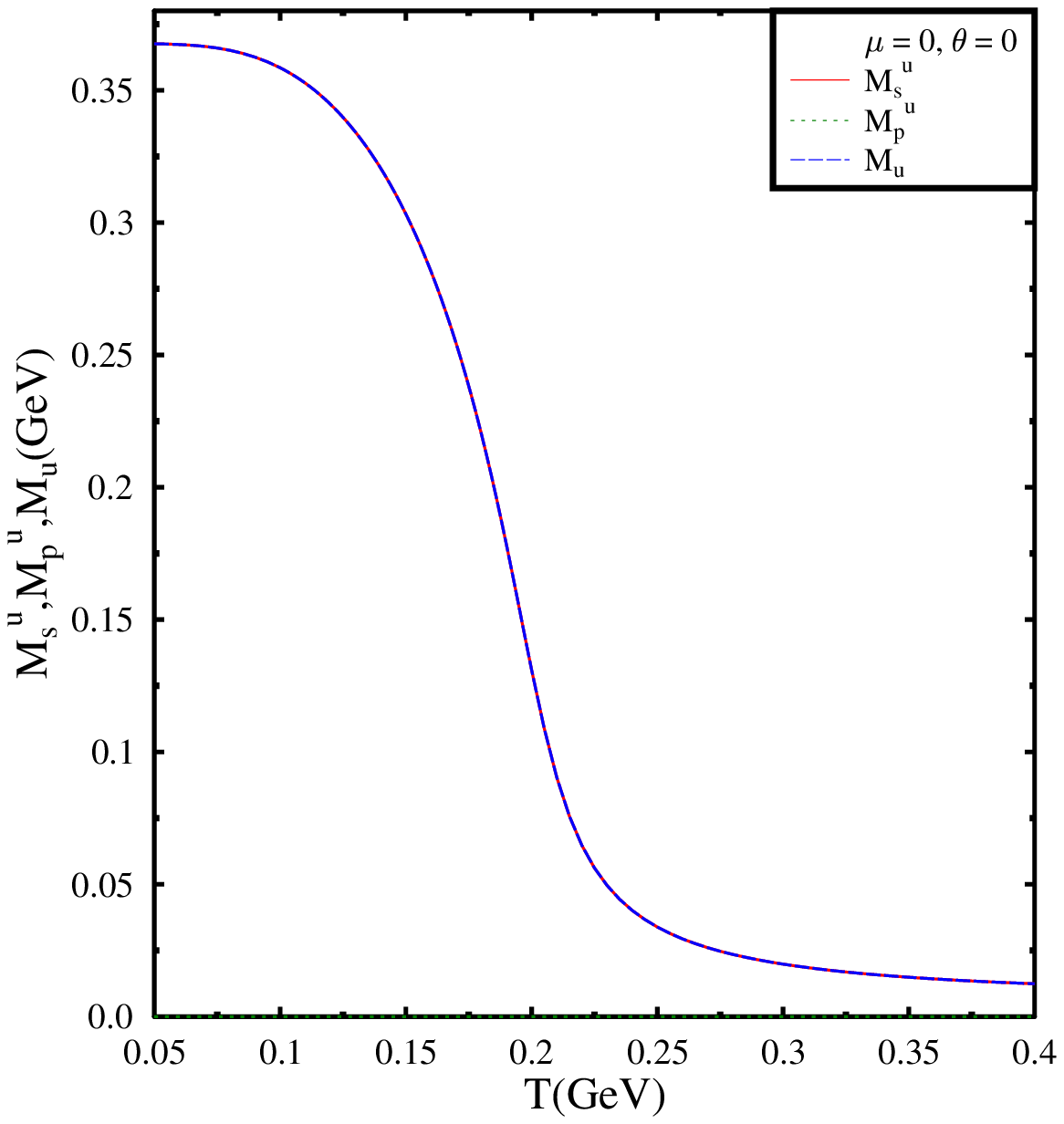}&
\includegraphics[width=6cm,height=6cm]{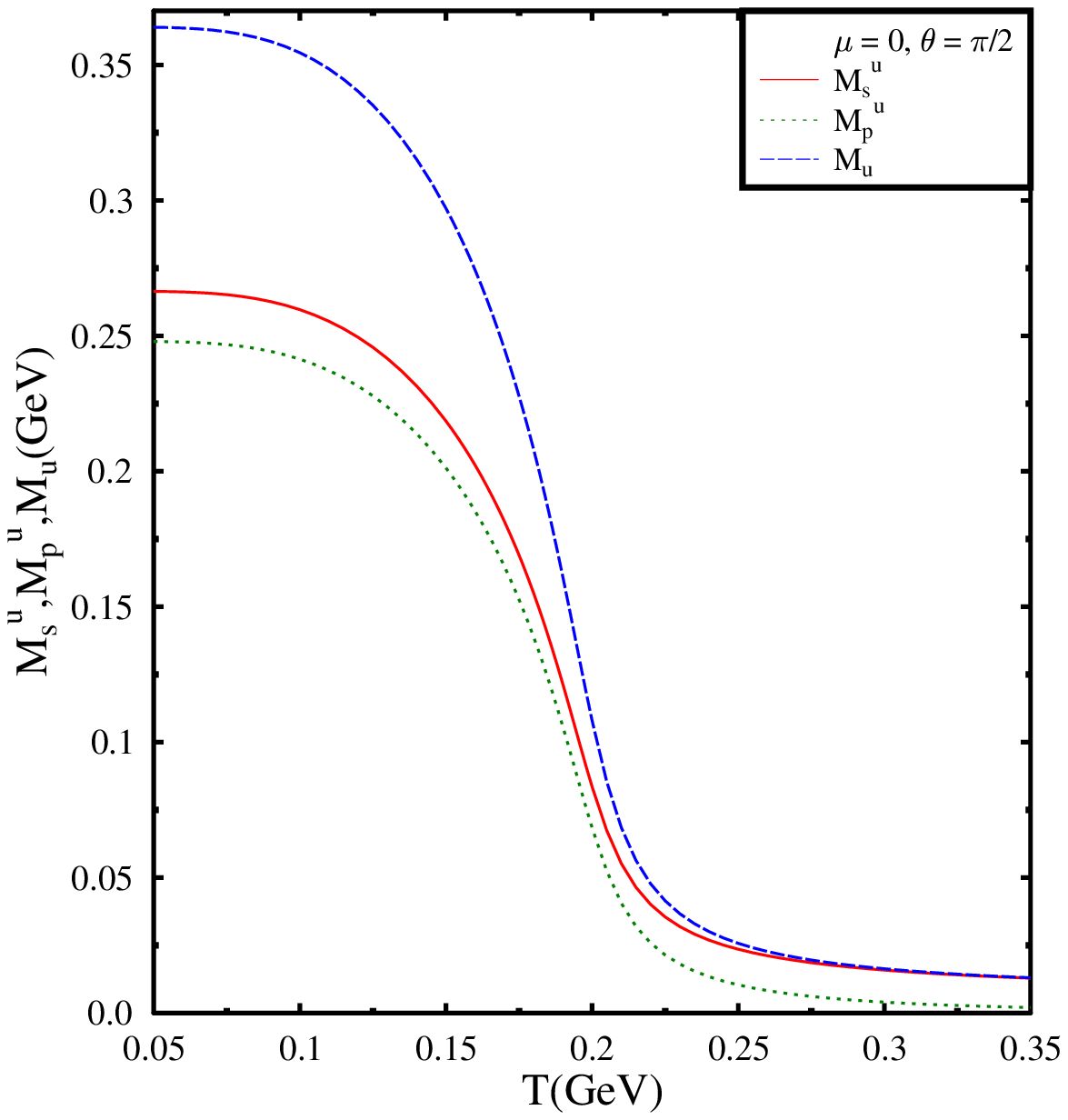}&
\includegraphics[width=6cm,height=6cm]{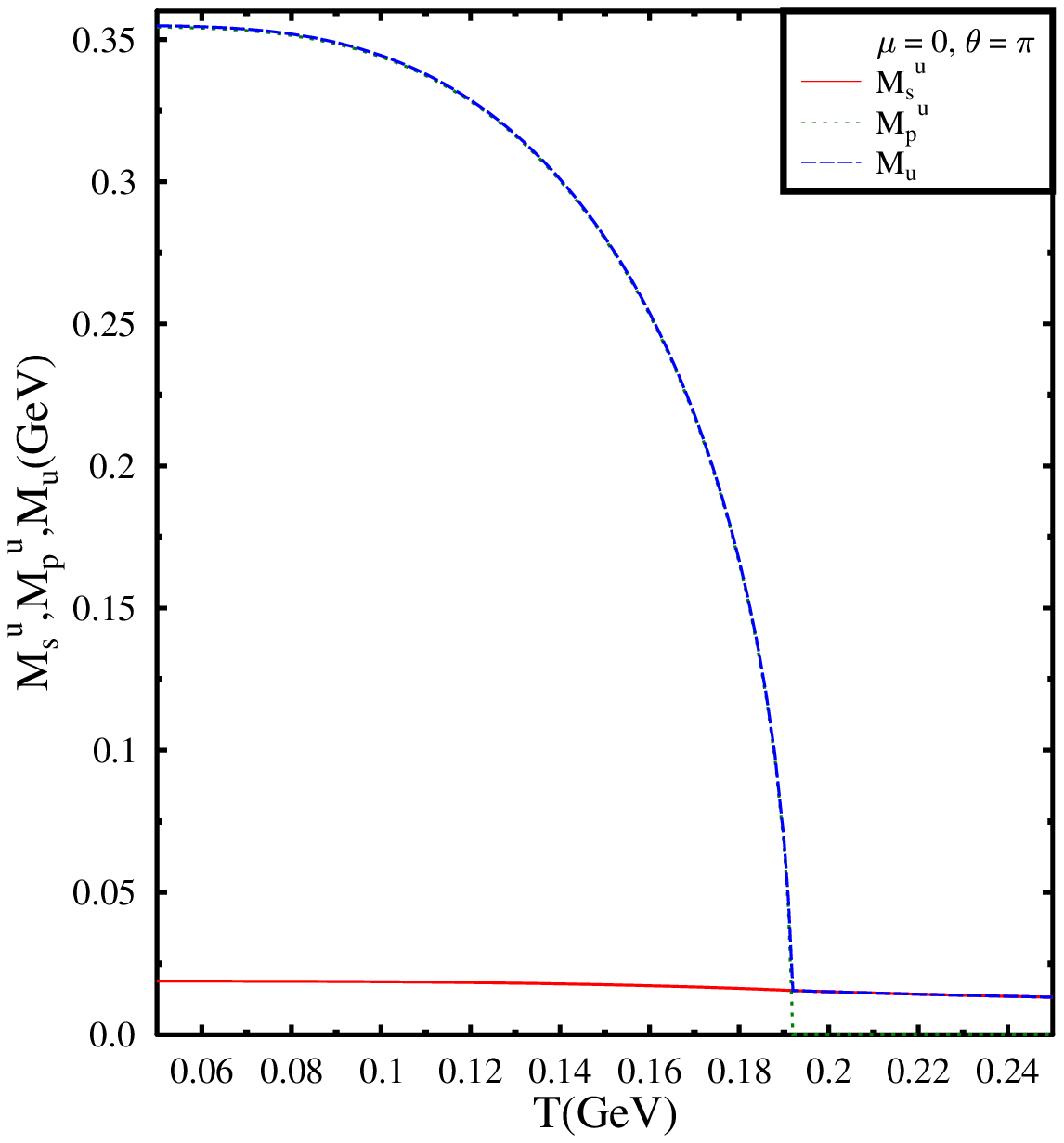}\\
Fig. 3-a & Fig. 3-b & Fig. 3-c
\end{tabular}
\end{center}
\caption{Quark masses as a function of temperature at zero quark chemical potential
for $\theta=0$ (Fig.a), $\theta=\pi/2$(Fig b) and $\theta=\pi$.}
\label{fig3}
\end{figure}

\begin{figure}[h]
\vspace{-0.4cm}
\begin{center}
\begin{tabular}{c c}
\includegraphics[width=9cm,height=9cm]{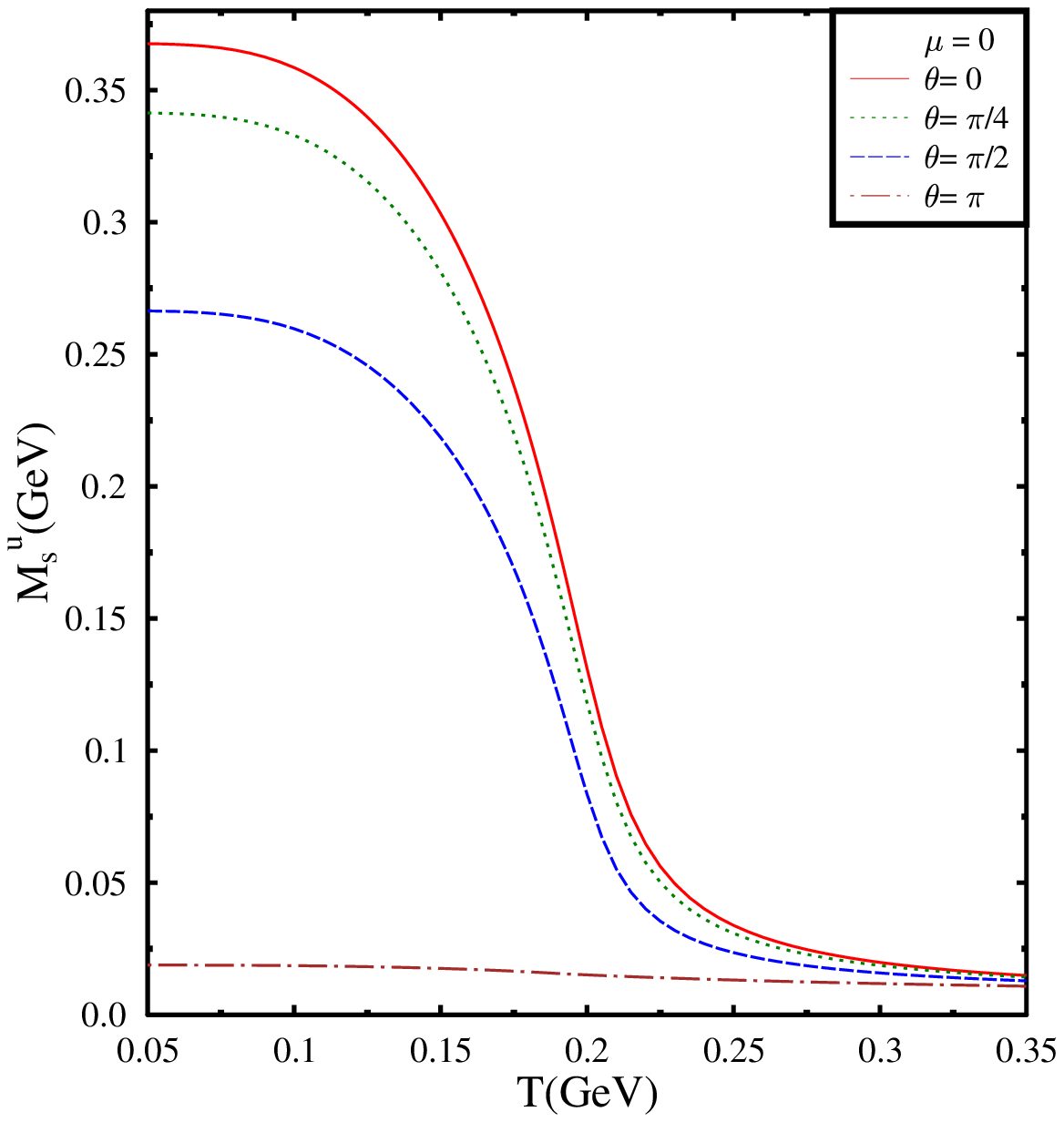}&
\includegraphics[width=9cm,height=9cm]{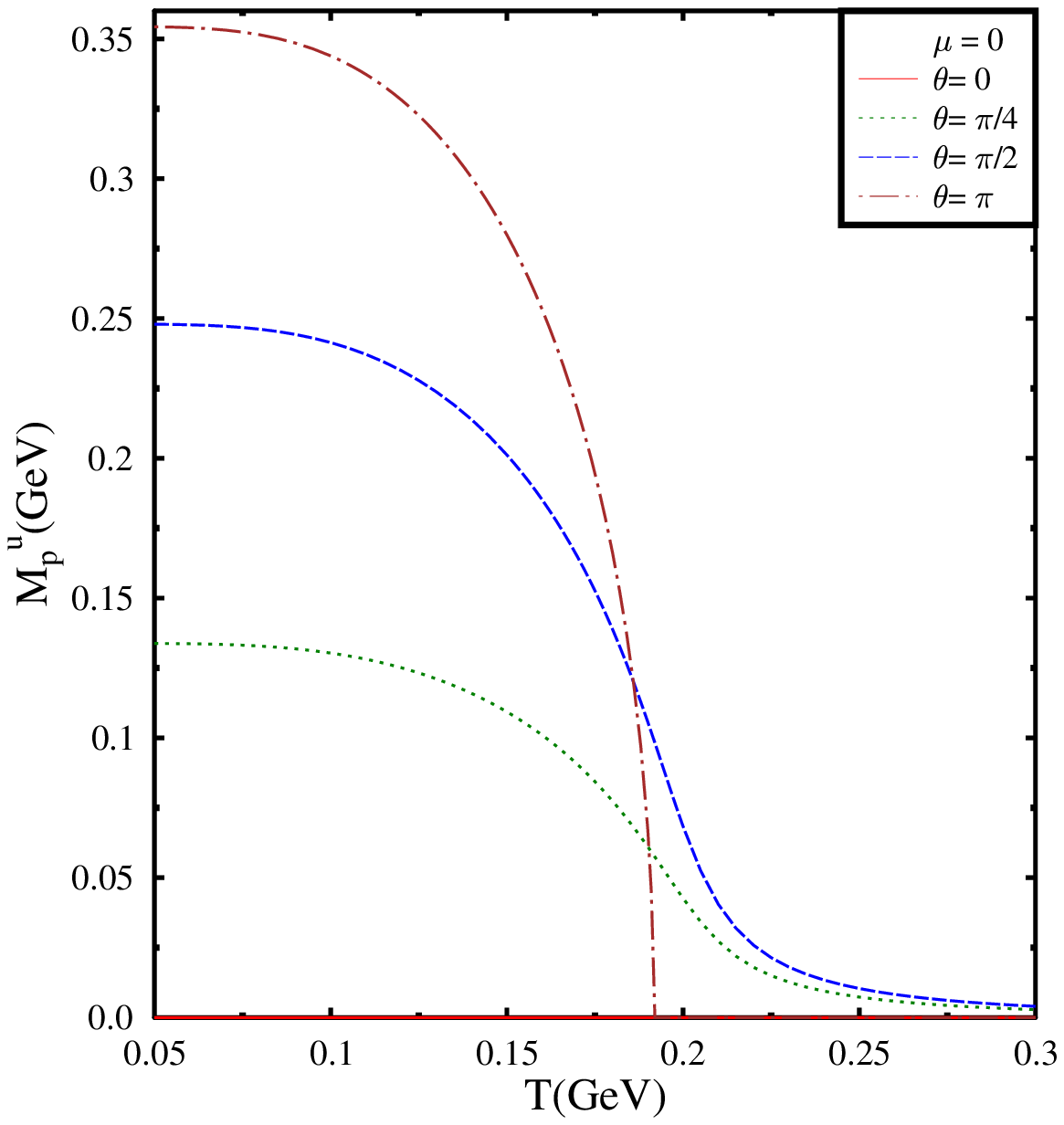}\\
Fig. 4-a & Fig. 4-b
\end{tabular}
\end{center}
\caption{The nature of transition of the scalar (Fig.a) and pseudoscalar (Fig.b) condensates with 
temperature at zero baryon density for different values of $\theta$.}
\label{fig4}
\end{figure}

\begin{figure}[htbp]
\vspace{-0.4cm}
\includegraphics[width=8cm,height=8cm]{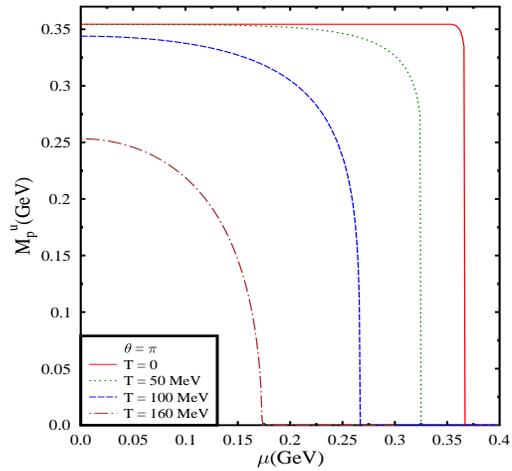}
\caption{Contribution from the pseudoscalar condensates to the u-quark mass 
as a function of quark chemical potential for different temperatures.
Here we have taken  $\theta=\pi$.}
\label{fig5}
\end{figure}

The CP restoring transition in the present 3-flavor NJL model turns to be second order for zero 
chemical potential case similar to the results for the 2-flavor case \cite{bbtwo} unlike the case 
of linear sigma model coupled to quarks \cite{fragalsm}. The reason behind such different 
behavior regarding the order of the transition is due to the non analytic vacuum term in the NJL 
model \cite{bbtwo}. However, a first order CP transition is observed with finite chemical potential 
and small temperature. This is clearly shown in Fig.(\ref{fig5}) where we have shown the dynamical 
mass, $M_p^u$ arising from the pseudoscalar condensate, for different temperatures as a function 
of the quark chemical potential for $\theta=\pi$. While at zero temperature, the order parameter 
decreases discontinuously, as the temperature increases, it becomes less sharp and finally results 
in a second order transition at high temperature.

\begin{figure}[htbp]
\vspace{-0.4cm}
\includegraphics[width=8cm,height=8cm]{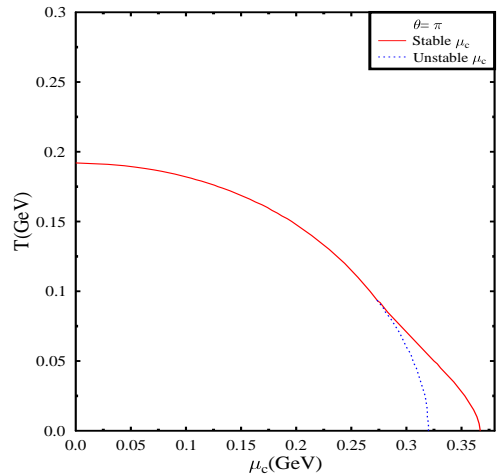}
\caption{The phase diagram for CP transition in the $T-\mu$ plane. 
The region between the solid line and the dotted line have 
solutions to the pseudoscalar mass gap equation but with higher 
thermodynamic potential. Here we have taken  $\theta=\pi$.}
\label{fig6}
\end{figure}

In Fig.(\ref{fig6}), we show the phase diagram in the plane of quark chemical potential and 
temperature for the CP violating transition. Since the transition is first order at zero 
temperature and second order at zero chemical potential, there is a tricritical point for this 
transition in this plane. This turns out to be ($\mu_c,T_c$)=(273,94) MeV. Including Polyakov loop 
for the two flavor NJL model, such a tricrtical point occurs at (209,165) MeV \cite{sakai}. First 
order transitions are associated with existence of meta-stable states. CP is restored in these 
meta stable states and these are the nontrivial solutions of the gap equation Eq.(\ref{psmass}), 
however with lower pressure than the stable solutions. In the phase diagram of Fig.(\ref{fig6}), 
such solution exist in the region between solid line and the dotted line.

\section{summary}
To summarize, we have tried to examine the effect of $\theta$ vacuum on the phase diagram of 
strong interaction. This is investigated within the framework of a 3-flavor NJL model. The effect 
of CP violating $\theta$ term of QCD is incorporated through the KMT determinant interaction term 
in the quark space.

In the two flavor NJL model, it has been observed that spontaneous CP violation for $\theta=\pi$ 
occurs depending on the  magnitude of current quark masses as well as the strength of the 
determinant coupling \cite{bbone}. In the present case of 3-flavor NJL model, we observe that 
spontaneous CP violation occurs for $\theta=\pi$ for the phenomenologically consistent parameters 
\cite{rehberg} for  the current quark masses as well as the strength of the determinant coupling.

To calculate the thermodynamical potential in presence of a $\theta$ term, instead of performing a 
chiral rotation of the quark field operators \cite{bbone,sakai}, we have used a variational 
approach using an explicit construct for the ground state. We have considered an ansatz state 
general enough to have condensates both in scalar as well as pseudoscalar channel. The ansatz 
functions are determined through minimization of the thermodynamic potential.

Apart from the temperature that has been considered earlier for two flavor cases 
\cite{bbtwo,fragalsm} we have also considered the effect of finite quark chemical potential and 
discussed the phase diagram in the $T-\mu$ plane. It turns out that for the range of temperature 
and the chemical potentials we have considered, the strange quark condensates do not get 
dynamically generated in the CP violating pseudoscalar channel even for non vanishing $\theta$. 
None the less the strange quark antiquark condensates in the scalar channel do affect the 
pseudoscalar light quark condensates through the flavor mixing coupled gap equations.

The CP restoring transition turns out to be a crossover for non vanishing values of $\theta$ that 
become a second order transition for $\theta=\pi$ for zero chemical potential. However, such a 
transition is a first order transition at small enough temperature and as the quark chemical 
potential increases. This leads to a tricritical point in the the phase diagram for CP transition.

\def\endm{C. Baker et al., {\PRL{97}{131801}{2006}};
 J. Kim and G. Carosi, Rev. Mod. Phys. {\bf 82}, 557 (2010).}
\def\pecceiquinn{R. D. Peccei and H. R. Quinn, {\PRL{38}{1440}{1977}}; {\PRD{16}{1791}{1977}}.}
\def\vafawit{C. Vafa and E. Witten, {\PRL{53}{535}{1984}}.}
\def\dashen{R. Dashen, {\PRD{3}{1879}{1971}}}
\def\chpt{P. Vecchia and G. Veneziano, {\NPB{171}{253}{1980}};
 A. Smilga, {\PRD{59}{114021}{1999}}; M. Tytgat, {\PRD{61}{114009}{2000}};
 G. Akemann, J. Lenaghan and K. Splittorff, {\PRD{65}{085015}{2002}};
 M. Creutz, {\PRL{92}{201601}{2004}};
 M. Metlitski and A. Zhitnitsky, {\NPB{731}{309}{2005}}; {\PLB{633}{721}{2006}}.}
\def\fragalsm{A. Mizher and E. Fraga, {\NPA{820}{247c}{2009}}; {\NPA{831}{91}{2009}}.}
\def\cpnjl{T. Fujihara, T. Inagaki and D. Kimura, {\PTP{117}{139}{2007}}.}
\def\bbone{D. Boer and J. Boomsma, {\PRD{78}{054027}{2008}}.}
\def\bbtwo{D. Boer and J. Boomsma, {\PRD{80}{034019}{2009}}.}
\def\sakai{Y. Sakai, H. Kouno, T. Sasaki and M. Yahiro, {\PLB{705}{349}{2011}}.}
\def\dimacp{D. Kharzeev, Annals Phys. {\bf 325}, 205 (2010).}
\def\cme{D. Kharzeev, {\PLB{633}{260}{2006}};
 D. Kharzeev, L. McLerran and H. Warringa, {\NPA{803}{227}{2008}};
 K. Fukushima, D. Kharzeev and H. Warringa, {\PRD{78}{074003}{2008}};
 K. Fukushima, M. Ruggieri and R. Gatto, {\PRD{81}{114031}{2010}}.}
\def\starexp{B. Abelev et al. [STAR Collaboration], {\PRL{103}{251601}{2009}};
 {\PRC{81}{054908}{2010}}.}
\def\klevansky{S. Klevansky, Rev. Mod. Phys. {\bf 64}, 649 (1992).}
\def\rehberg{P. Rehberg, S. P. Klevansky and J. Huefner, {\PRC{53}{410}{1996}}.}
\def\amspm{A. Mishra and S. P. Misra, {\ZPC{58}{325}{1993}}.}
\def\hmspmnjl{H. Mishra and S. P. Misra, {\PRD{48}{5376}{1993}}.}
\def\tfd{H. Umezawa, H. Matsumoto and M. Tachiki {\it Thermofield dynamics
and condensed states} (North Holland, Amsterdam, 1982);
P. A. Henning, Phys. Rep. {\bf 253}, 235 (1995).}
\def\amph4{A. Mishra and H. Mishra, {\JPG{23}{143}{1997}}.}
\def\hmcsc{A. Mishra and H. Mishra, {\PRD{74}{054024}{2006}}.}
\def\bccsb{B. Chatterjee, A. Mishra, H. Mishra, {\PRD{84}{014016}{2011}}.}
\def\grosspisarski{D. Gross, R. Pisarski and L. Yaffe, Rev. Mod. Phys. {\bf 53}, 43 (1981).}

\end{document}